\documentclass[10pt, conference]{IEEEtran}
\IEEEoverridecommandlockouts
\usepackage{amsmath}

\usepackage{amssymb}
\usepackage{mathtools}
\usepackage{multirow}
\usepackage{xcolor}
\usepackage{colortbl}

\def\BibTeX{{\rm B\kern-.05em{\sc i\kern-.025em b}\kern-.08em
    T\kern-.1667em\lower.7ex\hbox{E}\kern-.125emX}}
    
\begin{document}
\title{Optimizing Neural Networks with Learnable Non-Linear Activation Functions via Lookup-Based FPGA Acceleration}
% \author{
%     \IEEEauthorblockN{Mengyuan Yin\textsuperscript{*}}
%     \IEEEauthorblockA{\textit{Institute of High Performance Computing} \\
%                       \textit{A*STAR, Singapore}\\
%                       yin\_mengyuan@ihpc.a-star.edu.sg}
%     \and
%     \IEEEauthorblockN{Benjamin Chen Ming Choong\textsuperscript{*}}
%     \IEEEauthorblockA{\textit{Institute of High Performance Computing} \\
%                       \textit{A*STAR, Singapore}\\
%                       benjamin\_choong\_chen\_ming@ihpc.a-star.edu.sg}
%     \and
%     \IEEEauthorblockN{Chuping Qu}
%     \IEEEauthorblockA{\textit{Institute of High Performance Computing} \\
%                       \textit{A*STAR, Singapore}\\
%                       qu\_chuping@ihpc.a-star.edu.sg}
%     \and
%     \IEEEauthorblockN{Rick Siow Mong Goh}
%     \IEEEauthorblockA{\textit{Institute of High Performance Computing} \\
%                       \textit{A*STAR, Singapore}\\
%                       gohsm@ihpc.a-star.edu.sg}
%     \and
%     \IEEEauthorblockN{Weng-Fai Wong}
%     \IEEEauthorblockA{\textit{National University of Singapore} \\
%                       \textit{Singapore}\\
%                       wongwf@comp.nus.edu.sg}
%     \and
%     \IEEEauthorblockN{Tao Luo}
%     \IEEEauthorblockA{\textit{Institute of High Performance Computing} \\
%                       \textit{A*STAR, Singapore}\\
%                       luo\_tao@ihpc.a-star.edu.sg}
% }

\author{
    \IEEEauthorblockN{Mengyuan Yin\textsuperscript{1*},
                      Benjamin Chen Ming Choong\textsuperscript{1*},
                      Chuping Qu\textsuperscript{1}}
    \IEEEauthorblockN{Rick Siow Mong Goh\textsuperscript{1},
                      Weng-Fai Wong\textsuperscript{2},
                      Tao Luo\textsuperscript{1\dag}}
    \IEEEauthorblockA{\textit{Institute of High Performance Computing, A*STAR, Singapore}}
    \IEEEauthorblockA{\textit{National University of Singapore, Singapore}}
    \thanks{\textsuperscript{*}Mengyuan Yin and Benjamin Chen Ming Choong contributed equally to this work.}
    \thanks{\textsuperscript{\dag}Tao Luo is the corresponding author to this work.}
}

\maketitle

\renewcommand{\thefootnote}{}
\footnotetext{This is the authors’ preprint version of a paper accepted for publication in the International Conference on Computer-Aided Design (ICCAD) 2025.}
\renewcommand{\thefootnote}{\arabic{footnote}}

\begin{abstract}
% Machine learning models with learnable non-linear activation functions have recently gained increasing attention. 
% Emerging models such as Kolmogorov-Arnold Networks (KANs) have demonstrated potential to outperform conventional models with fixed activation functions in terms of size, accuracy, and interpretability, particularly in small-scale AI-for-science tasks. 
%Machine learning models with learnable non-linear activation functions, like Kolmogorov-Arnold Networks (KANs), have shown potential to outperform conventional models with fixed activations in size, accuracy, and interpretability, especially for small-scale AI-for-science tasks.
%Machine learning models that utilize learnable nonlinear activation functions, such as Kolmogorov-Arnold Networks (KANs), have demonstrated considerable potential in surpassing traditional models with fixed activation functions. These improvements are particularly notable in key metrics such as model size, accuracy, and interpretability, especially in the context of small-scale AI-for-science applications. 
Learned activation functions in models like Kolmogorov-Arnold Networks (KANs) outperform fixed-activation architectures in terms of accuracy and interpretability; however, their computational complexity poses critical challenges for energy-constrained edge AI deployments. Conventional CPUs/GPUs incur prohibitive latency and power costs when evaluating higher order activations, limiting deployability under ultra-tight energy budgets.
We address this via a reconfigurable lookup architecture with edge FPGAs. By coupling fine-grained quantization with adaptive lookup tables, our design minimizes energy-intensive arithmetic operations while preserving activation fidelity. FPGA reconfigurability enables dynamic hardware specialization for learned functions, a key advantage for edge systems that require post-deployment adaptability.
% Furthermore, we minimize resource utilization and energy through a fine-grained implementation of each function on the reconfigurable fabric of FPGAs. 
% The hardware design is fully automated and fine-grained to support different quantization levels and model shapes on the reconfigurable fabric of FPGAs.
%
%In experiments on KANs (in which unique activation functions dominate), our hardware design outperforms edge CPUs and GPUs in computation time and energy efficiency while maintaining the same level of accuracy and a small resource footprint.
%Experimental evaluations on KANs—where unique activation functions play a critical role—demonstrate that our FPGA-based design achieves superior computational speed and more than 10× improvement in energy efficiency compared to edge CPUs and GPUs. Notably, these performance gains are achieved without compromising model accuracy or significantly increasing resource utilization, ensuring a compact and efficient hardware footprint.
Evaluations using KANs - where unique activation functions play a critical role—demonstrate that our FPGA-based design achieves superior computational speed and over $10^4$ times higher energy efficiency compared to edge CPUs and GPUs, while maintaining matching accuracy and minimal footprint overhead. This breakthrough positions our approach as a practical enabler for energy-critical edge AI, where computational intensity and power constraints traditionally preclude the use of adaptive activation networks.
%Experimental results demonstrate that our design outperforms both edge CPUs and GPUs.

\end{abstract}


\section{Introduction}
\label{sec:intro}

% Learnable activation functions
The development of effective activation functions has long been a central focus in machine learning research to enhance neural network capabilities. 
%Recent work has increasingly shifted toward trainable activation functions as alternatives to traditional fixed operations like ReLU~\cite{relu} and Leaky ReLU~\cite{leaky_relu}, driven by their potential to improve model expressivity and task-specific adaptability.
Neural networks with trainable activation functions represent an important and actively explored class of models, attracting growing research interest due to their potential to enhance model expressivity and adaptability to specific tasks~\cite{dnn_trainable_act} - complementing models with traditional fixed functions such as ReLU~\cite{relu} and Leaky ReLU~\cite{leaky_relu}.
Learnable activation functions can be classified into two main categories: parameterized standard activation functions and ensemble-based activation functions~\cite{trainable_act_func_survey}. 
The first category extends traditional fixed-shape activation functions by introducing trainable parameters. However, these functions remain similar to their non-trainable counterparts, often resulting in only modest improvements in flexibility and expressiveness~\cite{trainable_act_func_survey}.
The second category combines a set of basis functions, which can either be fixed-shape or trainable. These basis functions can work together to capture complex patterns in the data. Examples of ensemble-based functions include S-shaped ReLU~\cite{s_relu} and adaptive activation functions~\cite{adapt_act_func}.

% KAN
Kolmogorov–Arnold Networks (KANs)~\cite{kan_liu} were recently proposed as an alternative to Multi-Layer Perceptrons (MLPs). 
They feature fully connected layers with ensemble-based learnable activation functions applied to edges. 
Research has demonstrated that KANs can achieve comparable accuracy with a smaller computational graph than MLPs and offer enhanced interpretability~\cite{kan_liu}.
This characteristic makes it well-suited for AI + Science applications~\cite{kan_liu_2}.
The promising results with KANs highlight the potential of machine learning models featuring learnable activation functions for effective pattern mapping, paving the way for further development and broader applications in the field.

% FPGA
However, since KAN activation functions are higher-order and distinct, inference on CPUs and GPUs is resource and energy-intensive.
The multiple operations required by higher-order functions increase computational demand, and the complexity and variability in activation functions disrupt the memory access efficiency that CPUs and GPUs are optimized for in handling uniform tasks~\cite{ma_bert}. 
% \textcolor{red}{[here can cite the MA-BERT work]}
FPGAs, on the other hand, allow for bit-level reconfigurability, enabling resource and energy savings through quantization~\cite{finn}. 
% \textcolor{red}{[here can cite a work on FPGA for quantised NN]}
Additionally, FPGAs can be customized for parallelism and pipelining, and hence reduces computation time.
Therefore, in this paper, we develop an acceleration platform on FPGA that is adaptable to neural networks with learnable activation functions.

% LUT 
Various methods can be employed to accelerate this class of neural networks on FPGAs.
The most direct approach is to program the general-purpose FPGA logic to perform arithmetic operations.  
While this offers flexibility, implementing complex arithmetic directly using look-up tables (LUTs) and flip-flops often results in inefficient designs with high latency and increased power consumption~\cite{fpga}.

Alternatively, dedicated digital signal processing (DSP) slices - which support high-performance multiply accumulate (MAC) operations - can be used to accelerate neural computations~\cite{dsp_acc}.  
However, because the learned activation functions are different and require unique operations, configuring and instantiating DSP slices to accommodate these variations is challenging.

Another approach involves in-memory computation using LUTs. 
Research has shown that this method is effective in accelerating quantized neural networks~\cite{tlmac}.
Inspired by this, our accelerator utilizes lookup-based architecture. 
By programming the mappings of learned activation functions into LUT initializations, our design enables single-cycle access to precomputed activation function outputs and supports a variety of learned activation functions.

% \textcolor{blue}{
% Machine learning models with learnable activation functions are often highly complex and involve higher-order computations. For example, in the Kolmogorov–Arnold Network (KAN) model, each activation function is implemented using a spline function, as described in Equation~\ref{eq:spline_definition}. This spline function comprises $G + k$ nonlinear B-spline basis functions, where $G$ and $k$ are specific parameters of the model. The use of these basis functions increases the complexity of the activation functions, making the process of programming FPGA logic to dynamically evaluate each activation function both time-consuming and energy-intensive.}

% \textcolor{blue}{Moreover, the activation functions differ both within a single KAN model and across different trained models, making direct programming overly specific and limiting the architecture’s generalizability. 
% To address this, lookup-based computing is considered to accelerate this kind of neural networks.}

% QUANT
The lookup-based implementation requires the quantization of activation functions, which presents two main challenges. First, minimizing the number of bits is essential for scalability. For a large KAN model to be deployable on an FPGA board, the bit width must be kept low. Second, achieving high accuracy is difficult, as the constraints of low-bit quantization, imposed to meet resource limitations, can significantly reduce accuracy compared to the original non-quantized model.

% Contribution
The contributions of this paper are as follows:
\begin{itemize}
    \item Propose a novel lookup-based architecture to accelerate neural networks with learnable activation functions.
    \item Introduce an efficient non-uniform quantization scheme that minimizes computational load during inference while maintaining accuracy.
    \item Demonstrate that our FPGA-based design outperforms edge CPUs and GPUs using the KAN model as a case study.
\end{itemize}

% Overview
In this paper, Section~\ref{sec: background} reviews the structure of KANs. 
Section~\ref{sec: sw} provides a detailed explanation of our proposed quantization scheme and the considerations behind it.
Section~\ref{sec: hw} describes the designed hardware architecture and its implementation details.
Section~\ref{sec: result} presents evaluations using the spherical harmonic function regression, and the MNIST dataset to demonstrate that our proposed acceleration platform for KANs outperforms edge CPUs and GPUs in computation time and energy efficiency, while maintaining accuracy and a compact resource footprint.
\section{Background and Notation}
\label{sec: background}

% KAN
\subsection{Neural Networks with Learnable Activation Functions}
Unlike neural networks with fixed activation functions, those equipped with trainable activation functions introduce more representability and flexibility into each edge connecting neurons. This enables the network to better capture complex relationships within the data with fewer neurons. 

An example of such an architecture is the Kolmogorov–Arnold Network (KAN). KAN is a fully connected neural network whose activation functions are inspired by the Kolmogorov–Arnold representation theorem~\cite{kan_liu}. A KAN model with \(L\) layers and an input vector \(\mathbf{x} \in \mathbb{R}^{n^0}\) can be expressed as:
\begin{equation}
\text{KAN}(\mathbf{x}) = \left( \Phi_{L-1} \circ \Phi_{L-2} \circ \cdots \circ \Phi_1 \circ \Phi_0 \right) \mathbf{x},
\label{eq:kan_definition}
\end{equation}
where each \(\Phi_l\) with \(1 \leq l \leq L\) represents a layer-specific transformation comprising learnable parameters and activation functions. Since the KAN model is fully connected, within each layer \(l\), the output of the \(j\)-th neuron, denoted as \(x^l_j\), is given by:
\begin{align}
y^l_{i \rightarrow j} &=\phi^{l-1}_{i \rightarrow j}\left( x^{l-1}_i \right) \nonumber\\ 
x^l_j &= \sum_{i=1}^{n_{l-1}} y^l_{i \rightarrow j}, \quad j = 1, \dots, n^l
\label{eq:layer_definition}
\end{align}
where \(n^l\) is the number of neurons in layer \(l\), and \(\phi^{l-1}_{i \rightarrow j}\) contains the learnable activation function and weights applied to the \(i\)-th input neuron from the previous layer. Every \(\phi\) contains two parts, a Sigmoid Linear Unit (SiLU) residual function and a spline function: 
% \textcolor{red}{SiLU is not a well recognised term, so please give the full name when first use it}
\begin{equation}
\phi(x) = w_b \cdot \text{SiLU}(\mathbf{x}) + w_s \cdot \text{spline}(x),
\label{eq:phi_definition}
\end{equation}
where the spline function of order is a linear combination of B-splines.
\begin{equation}
\text{Spline}(x) = \sum_i^{G+k} c_i B_i(x)
\label{eq:spline_definition}
\end{equation}

\begin{figure}[h!]
    \centering
    \includegraphics[width=0.3\textwidth]{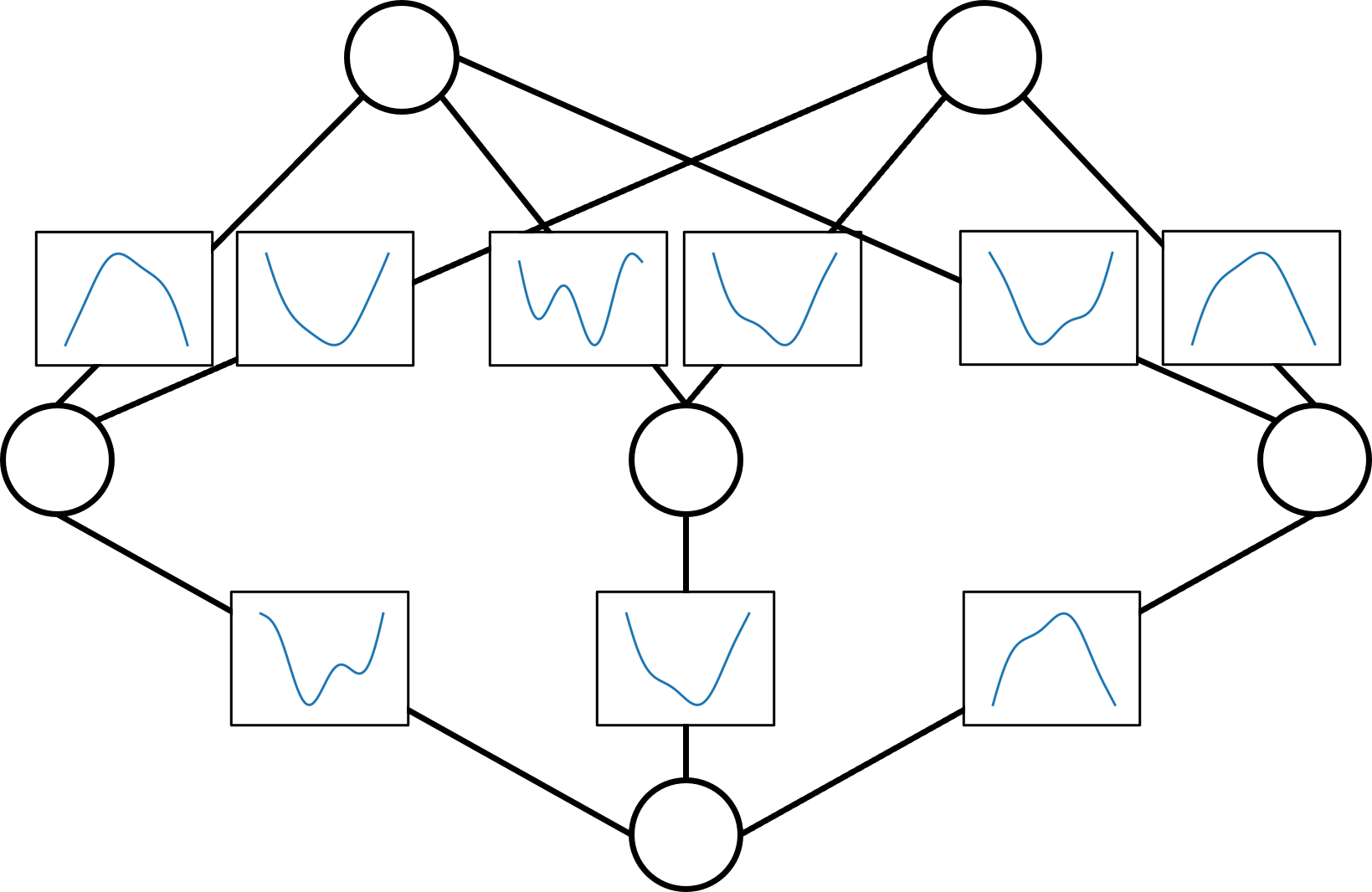}
    \caption{An illustration of a KAN model with shape [2,3,1]}
    \label{fig:kan}
\end{figure}

Figure~\ref{fig:kan} shows an instance of the KAN model. Assume that the number of neurons per layer is \(n^0 = n^1 = \dots = N\), and each spline function is of order \(k\) with \(G\) intervals (resulting in \(G + 1\) grid points). The total number of parameters in the network is \(O(N^2L(G + k))\). This parameter count exceeds that of an equivalently shaped MLP, which requires \(O(N^2L)\) parameters. However, KANs typically achieve comparable data fitting with much smaller \(N\) than MLPs. Rather than directly comparing KANs with MLPs, this work appreciates the value of learnable activation functions, which enable the network to effectively represent complex nonlinear relationships without the need for deep or wide architectures. Motivated by this advantage, we aim to develop an acceleration tool tailored for networks with learnable activation functions like KANs, addressing their computational demands and facilitating more energy-saving deployment.

\subsection{Table Lookup-based Hardware Accelerators}

Lookup-based computing has been employed to accelerate computation that is not efficiently supported on conventional hardware, such as CPUs and GPUs.
Examples in deep learning include non-uniform quantization \cite{cardinaux2020iteratively,wang2022learnable,guo2024fast}, where LUTs are used to implement the quantization function as a codebook, improving the representational capability of low-bit activations fetched from memory over uniform quantization.
In addition, for other complex functions including softmax \cite{vasyltsov2021efficient,chen2022approximate} and non-linear activations \cite{yu2022nn}, lookup-based computing is also used to complement conventional computing devices well-suited for linear arithmetic (e.g. GPUs), implementing complex operations including reciprocal exponentiation, square roots, GELU etc. with improved efficiency. 
In contrast to these works, our work targets the acceleration of arbitrary non-linear functions that are parameterized and learned during training.

Another class of works involves the use of lookup operations to replace the core computation of neural networks.
In \cite{wang2019lutnet,wang2022logic}, LUTs are used to implement the core XNOR operations in binary neural networks in the logic fabric of FPGAs.
Lookup-based neurons have also been used to optimize small models for ultra-low latency inference \cite{umuroglu2020logicnets} and embed the multivariate polynomial evaluation \cite{andronic2023polylut} in complex neuron functions.
For larger models such as convolutional neural networks \cite{tlmac} and transformer models \cite{park2022lut}, lookup-based computing has been employed to accelerate even matrix multiplication with sub-4-bit weights and activations.
Our proposed architecture is closest to these works by using lookup as the core computation of neural network inference, but targeting learned non-linear activation functions instead.

\section{Quantization of Learnable Activation Functions}
\label{sec: sw}
In lookup-based computing, complex non-linear functions are implemented using {\em look-up tables} (LUTs) as binary mappings from input to output bits.
Each $b_{\text{in}}$-input LUT is programmed to generate one output bit by storing $2^{b_{\text{in}}}$ bits covering all possible input combinations.
A collection of $b_{\text{out}}$ such LUTs can be combined to generate a $b_{\text{out}}$-bit output.
Hence, the number of $N$-input LUTs (also termed LUT-$N$) required for a non-linear function using LUTs is given by
\begin{equation} \label{eq:lut-scaling}
    \#\text{LUT-}N = b_{\text{out}} \cdot 2^{b_{\text{in}} - N}
\end{equation}
scaling linearly with output bits and exponentially with input bits.
However, this exponential scaling results in infeasible LUT costs for functions with high-precision inputs, such as 32-bit or 64-bit floating-point. 
To illustrate, an arbitrary logic function with 32-bit inputs would theoretically require 268 million LUT-4s for implementation, while the highest LUT-$4$ count of high-performance AMD Virtex Ultrascale+ FPGAs is at 9 million.

To address these scalability challenges, it is necessary to quantize the inputs and outputs of the learnable activation functions.
Let $y = \phi(x)$ be a learned activation function trained on data.
Furthermore, let $\bar{v}_{<b>}$ denote the $b$-bit quantization function on data $v$ in integer format, and $\hat{v}_{<b>}$ denote the corresponding quantized real value using a step size $s$.
As an example, using bit-width $b=5$ and step size $s=0.1$, $v = 2.18$ can be quantized as $\bar{v}_{<5b>} = 22 = (10110)_{2}$ and $\hat{v}_{<5b>} = 2.20$.

We apply uniform quantization on both inputs and outputs, given by
\begin{equation} \label{eq:uniform-quant}
    \bar{v} = \Bigl \lfloor \text{clamp}\Big( \frac{v-\min(v)}{s} \; , 0, \; 2^{b}-1 \Big) \Bigr \rceil
\end{equation}
where $\text{clamp}(\cdot)$ clips the values to the range between $0$ and $2^{b}-1$, and $\lfloor \cdot \rceil$ is the rounding function. 
The minimum value $\min(v)$ is subtracted to make $\bar{v}$ a non-negative integer.

We first present our method of \emph{quantizing activation functions} for scalable lookup - reducing the precision of inputs $x$ and outputs $y$.
Following this, we implement \emph{efficient precision conversion} to quantize data values between layers during run-time, requiring only fixed-point arithmetic instead of expensive floating-point dequantization and requantization. 

\subsection{Quantizing Activation Functions}
\label{sec: quant_act_func}
% \textcolor{red}{it is better to give a brief introduction of the following two-step method: Gloabal quantization and following fine-grained quantization}
In quantizing the inputs and outputs of learned activation functions, we first perform (1) \emph{global quantization}, where a single degree of precision is applied to all functions in a layer.
For further efficiency savings, we then apply (2) \emph{fine-grained quantization} to assign unique bit reductions to each function while maintaining accuracy.

\subsubsection{Global Quantization}
We first perform global quantization - quantizing all activation functions to an initial $b_{\text{in}}$-bit input precision and $b_{\text{out}}$-bit output precision while ensuring accuracy is preserved.
Step sizes $s$ are selected according to the neural network architecture, dividing the range of data values into $2^{b}$ equally-spaced levels.
Care should be taken to ensure activation functions of nodes/edges that must share the same data representation have compatible step sizes.

In KANs, each layer has been trained on a grid which clamps inputs within a fixed range of values.
This range is divided to obtain the input step size for each layer.
Conversely, output step sizes are determined from the output range of activation functions. 
As shown in Equation \ref{eq:layer_definition}, each output neuron \( x^{l}_{j} \) accumulates the results of all activation functions \( \phi^{l-1}_{i \rightarrow j}(x^{l-1}_i) \) of its input edges over index $i$. 
% , where \( i \) ranges from \( 1 \) to \( n_{l-1} \). Here, \( n_{l-1} \) represents the total number of neurons in layer \( l-1 \).
Hence, all activation functions of the same output neuron (same $j$) must share the same output step size, ensuring a shared data representation for efficient accumulation.

For global quantization, the scheme in Equation \ref{eq:uniform-quant} is applied, iteratively evaluating different bit-widths to yield a minimal $b$ for which a pre-determined error threshold is not exceeded.

\subsubsection{Fine-grained Quantization}
\label{sec: quant_fine_grain}
Upon training, we observe that not all learned activation functions require equal precision.
More aggressive fine-grained quantization can be applied as an additional optimization according to each activation function, translating into cost savings and potential speedup.

\begin{figure}[t]
\begin{center}
\centerline{\includegraphics[width=\columnwidth]{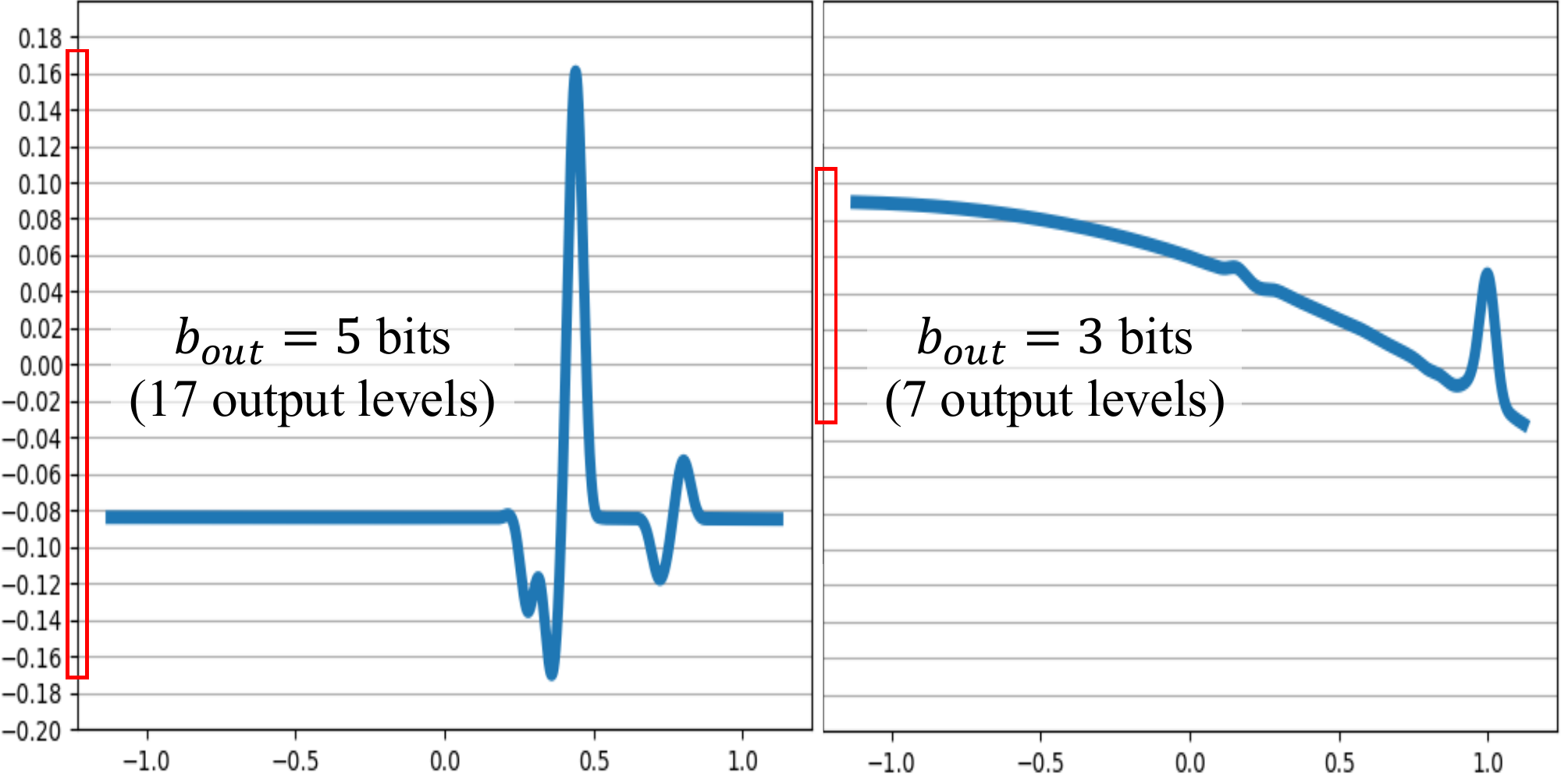}}
\caption{Fine-grained output quantization of two learned activation functions. The two functions have the same step size (y-axis resolution) but different ranges (red boxes), allowing different output bit-widths (5-bit for left function, 3-bit for right function) to be implemented on hardware.}
\label{fig:fine-grained-output-quant}
\end{center}
\end{figure}

Fine-grained quantization of activation function \emph{outputs} is relatively straightforward: while activation functions of the same output neuron share the same step size, each function can have very different output ranges.
Hence, an activation function with a smaller range is quantized to only the number of bits required to represent its range.
An example is illustrated in Figure \ref{fig:fine-grained-output-quant}, where the left function covers a larger output range and so requires more output bits (5-bit for 17 levels) compared to the right function (3-bit for 7 levels).

Using the same output step size $s$, the output bit-width of learned function $\phi^{l}_{i \rightarrow j}(\cdot)$ is given by 
\begin{equation}
    b^{l}_{\text{out}, i \rightarrow j} = \biggl \lceil \log_{2} \bigg( \frac{1}{s}(\max(\phi^{l}_{i \rightarrow j}(\cdot)) - \min(\phi^{l}_{i \rightarrow j}(\cdot))) \bigg) \biggr \rceil
\label{eq:bo}
\end{equation}
where $\lceil \cdot \rceil$ is the ceiling function.
To ensure correctness during accumulation, we first subtract $\min(\phi^{l}_{i \rightarrow j}(\cdot))$ as an offset, using an unsigned integer representation for the remaining values. 
In other words, the first level (binary integer $0$) represents real-valued offset $\min(\phi^{l}_{i \rightarrow j}(\cdot))$.
The offset subtraction must be applied to each edge after the activation function and before accumulating the contributions from all neurons in the previous layer to compute the value of a neuron in the next layer.
This fine-grained quantization of output is applied with no degradation from the global quantization step.
% \textcolor{red}{It is better to provide visual aids like figures with examples to demonstrate the Global quantization and fine-grained quantization.  BEN: will add figure here.}

The fine-grained quantization of activation function \emph{inputs} is arguably more impactful, due to the exponential scaling of LUTs with input bits, shown in Equation \ref{eq:lut-scaling}.
However, searching for a fine-grained input bit assignment for each function entails a vast combinatorial space that is challenging to search exhaustively.
Hence, we endeavor to identify functions with lower sensitivity to input quantization, targeting these functions for precision reduction.

To this end, we employ a heuristic to estimate the sensitivity of activation functions to input resolution.
For a given input $x$, the absolute output error from input quantization, denoted $\mathbf{E}_{q}(\phi)$ below, can be written as (with indices dropped for simplicity):
\begin{align}
    \mathbf{E}_{q}(\phi) &= \big| \phi(\hat{x}) - \phi(x)\big| \nonumber \\
    &= \big| \phi(x + \Delta x) - \phi(x) \big| \text{, where } \Delta x = \hat{x}-x
\end{align}
The value of $\Delta x$ varies according to the rounding of $x$ to the nearest quantized value.
However, as we increase $b$ towards infinite precision, $\Delta x$ becomes infinitely small and its variation negligible, allowing quantization error $\mathbf{E}_{q}(\phi)$ to be approximated as the \emph{absolute instantaneous gradient} of $\phi(x)$ at $x$.

\begin{figure}[t]
\begin{center}
\centerline{\includegraphics[width=\columnwidth]{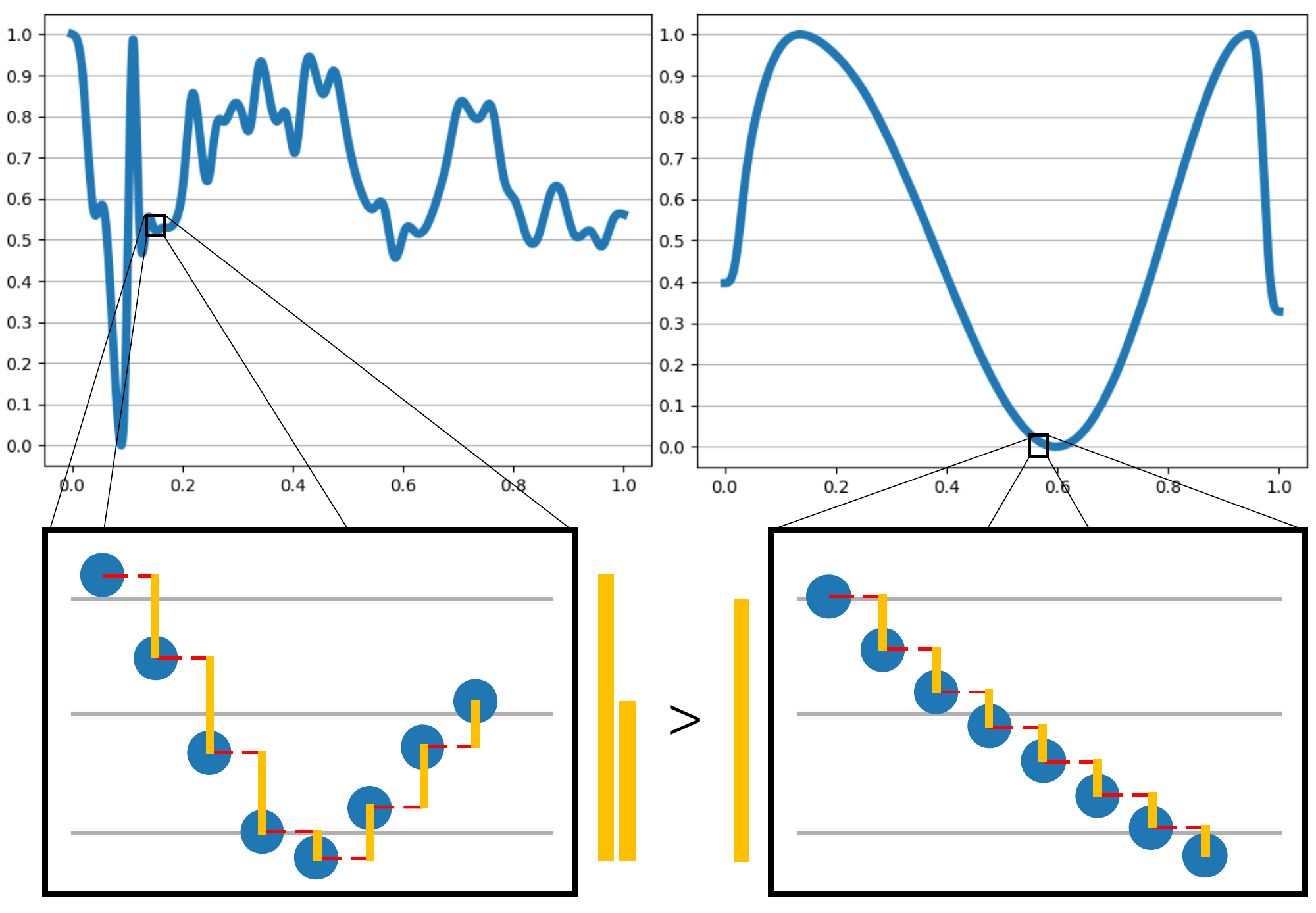}}
\caption{Our proposed ``sum of absolute differences" heuristic accumulates the absolute differences between neighboring values (orange bars) to estimate sensitivity to input resolution. This summation is performed across the entire function. Due to many steeper changes over $x$, the left function with $\mathbf{E}_{q}(\phi) = 6.06$ would typically require higher input resolution than the right function with $\mathbf{E}_{q}(\phi) = 3.27$.}
\label{fig:sum-of-differences}
\end{center}
\end{figure}

When discretized at high input resolution, the integration of absolute derivatives $\int |\phi'(x)| \, dx$ can be approximated as the \emph{sum of absolute differences} between neighboring values $x$ and $x^{+}$, given by
\begin{equation}
    \mathbf{E}_{q}(\phi) = \sum_{x} | \phi(x^{+}) - \phi(x)| 
\end{equation}
Empirically, we find this sum to be a useful estimate of the sensitivity of $\phi(x)$ to input resolution.
An example is illustrated in Figure \ref{fig:sum-of-differences}, comparing this ``sum of differences" heuristic between two functions with different quantization sensitivity.

For a scale-agnostic comparison of sensitivity, we first scale each activation function to have a range of 1. 
We then rank functions by estimated sensitivity and greedily quantize those with lower sensitivity first, while ensuring error/accuracy remains within a predefined threshold.
% For small KANs with few edges, each function can be quantized sequentially. 
% For larger KANs with thousands of edges, we first sort the functions in each layer by sensitivity, and iteratively reduce the precision of the set of functions with low sensitivity by 1 bit each time.
% We use binary search to partition the sorted functions into high- and low-sensitivity sets, ensuring the low-sensitivity set maintains within the error threshold upon quantization.
% \textcolor{red}{It is better to provide pseudo-code to better summaries the algorithm/methods proposed.}

\subsection{Efficient Precision Conversion}
\label{sec: quant_act_data}
Different activation functions often require different resolutions (step sizes) to maintain high precision.
Between layers, data must be quantized with the correct step size matching the input precision(s) of the quantized activation functions.
Conventional approaches often perform precision conversion through dequantization (conversion to full-precision) and requantization (division and rounding back to integer) \cite{jin2022f8net}; however, these operations incur expensive floating-point hardware.
Instead, we propose an efficient method to convert precisions using only fixed-point arithmetic, approximating the full-precision dequantization/requantization process using fixed-point scaling.

Our efficient precision conversion is best illustrated with an example.
In our illustration, we reuse the notation of the previous section, where $\bar{x}_{<b>}$ and $\hat{x}_{<b>}$ represent the $b$-bit quantized value of $x$ in integer and real value formats respectively.
We also drop neuron indices $i, j$ for simplicity.

Let $x^{l}$ denote the input to functions in layer $l$ that is received from the previous layer $l-1$.
Let layer $l$ be quantized to 4-bit input precision ($\hat{x}_{<4b>}$) with step size $s^{l}=0.5$.
Moreover, let the previous layer $l-1$ have 9-bit output precision ($\hat{x}_{<9b>}$) with step size $s^{l-1}=0.01$.

Consider the data value $x = 4.5$. Hence, the binary output of the preceding layer $\bar{x}_{<9b>}=4.5/0.01=450=(111000010)_{2}$, while the binary input for is $\bar{x}_{<4b>}=(4.5/0.5)=9=(1001)_{2}$.
Additional arithmetic is required to convert received output data $(101101)_{2}$ to compatible input $(1001)_{2}$.

To avoid the demand for floating-point operations, we explore the use of a low-bit fixed-point scaling factor $\bar{\alpha}$ to perform approximate format conversion, enabling the use of efficient fixed-point multiplication instead.
The conversion transformation is given in the following equations:
\begin{align}
    \hat{x}_{<b^{l}>} &= \hat{x}_{<b^{l-1}>} \nonumber \\
    \hat{x}_{<b^{l}>} \cdot s^{l} &= \hat{x}_{<b^{l-1}>} \cdot s^{l-1} \nonumber \\
    \hat{x}_{<b^{l}>} &= \hat{x}_{<b^{l-1}>} \cdot \alpha^{l} \text{, where } \alpha^{l} = \frac{s^{l-1}}{s^{l}}
\end{align}
The scaling factor $\alpha^{l}$ is then quantized to a fixed-point format, $\bar{\alpha}$, with a minimal number of significant bits such that accuracy is preserved.

Using the previous example, $\alpha=(0.01 / 0.5) = 0.02$.
We use $\bar{\alpha}^{l} = (0.00000101)_{2} = 0.01953 \approx 0.02$, where $\bar{\alpha}^{l}$ has eight fractional bits and three significant bits.
The minimum number of significant bits needed is found through iterative model evaluation with increasing bit count.
On hardware, this transformation becomes:
\begin{align}
    \bar{x}_{<b^{l}>} &= \bar{x}_{<b^{l-1}>} \cdot \bar{\alpha}^{l} \nonumber \\
    &= (111000010 \otimes 101 ) \texttt{ >> } 8 \nonumber \\
    &= 1000.11001010 \nonumber \\
    &\approx 1001 \text{ after rounding.}
\label{eq:precision_conv}
\end{align}
where $(111000010 \otimes 101 )$ is performed with 9-bit fixed-point multiplication, and $\texttt{ >> }$ represents the rightward bit shift operation.

Using a fixed-point $\bar{\alpha}$ thus allows more efficient hardware while minimizing error in format conversion.
This function is implemented in the quantization block in our hardware architecture.
\section{Hardware Architecture}
\label{sec: hw}

% Machine learning models with learnable activation functions are often highly complex and involve higher-order computations. For example, in the Kolmogorov–Arnold Network (KAN) model, each activation function is implemented using a spline function, as described in Equation 4. This spline function comprises $G + k$ nonlinear B-spline basis functions, where $G$ and $k$ are specific parameters of the model. The use of these basis functions increases the complexity of the activation functions, making the process of programming FPGA logic to dynamically evaluate each activation function both time-consuming and energy-intensive. 

% Moreover, the activation functions differ both within a single KAN model and across different trained models, making direct programming overly specific and limiting the architecture’s generalizability. 
% To address this, our architecture employs a lookup-based design. 

The hardware architecture of the lookup-based accelerator comprises three main components.
First, a quantization block performs input quantization. It quantizes the input for the model when $l = 1$, and ensures compatibility of the quantization precision level between layers when $1 < l \leq L$. This prepares the input for the subsequent activation function.
Next, a LUT pool implements the activation function $\phi(\cdot)$ by mapping the quantized input values to their corresponding activated outputs.
Both the quantization block and the LUT pool are instantiated per edge in the neural network.
At the end of each layer, an accumulator aggregates the outputs of all activation functions, as defined in Equation~\ref{eq:layer_definition}.
The accumulator is instantiated per output neuron of the layer.
% Additionally, to overcome the scalability challenges associated with full-precision bitwidth implementation, we incorporate a quantization model to fulfill the job as described in Section~\ref{sec: sw}. 
% By applying the quantization, we reduce the bitwidth of numerical representations, significantly lowering storage requirements and accelerating computations while maintaining an acceptable level of accuracy. 

% Each activation function \(\phi^l_{i\rightarrow j}\) in Equation~\ref{eq:phi_definition} is implemented using a quantization block and a LUT pool. 
% Here, \(l\) represents the layer index with \(1 \leq l \leq L\), where \(L\) is the total number of layers in the KAN model. 
% The indices \(i\) and \(j\) represent the positions of neurons within each layer, with \(1 \leq i \leq n_l\) and \(1 \leq j \leq n_{l+1}\), where \(n_l\) and \(n_{l+1}\) denote the number of neurons in the layers \(l\) and \(l+1\), respectively.

% Figure~\ref{fig:act_func} illustrates the hardware implementation of a single layer in the KAN model, where each \(x\) represents a neuron and Q denotes a quantization block. 

\begin{figure}[t]
    \centering
    \includegraphics[width=\columnwidth]{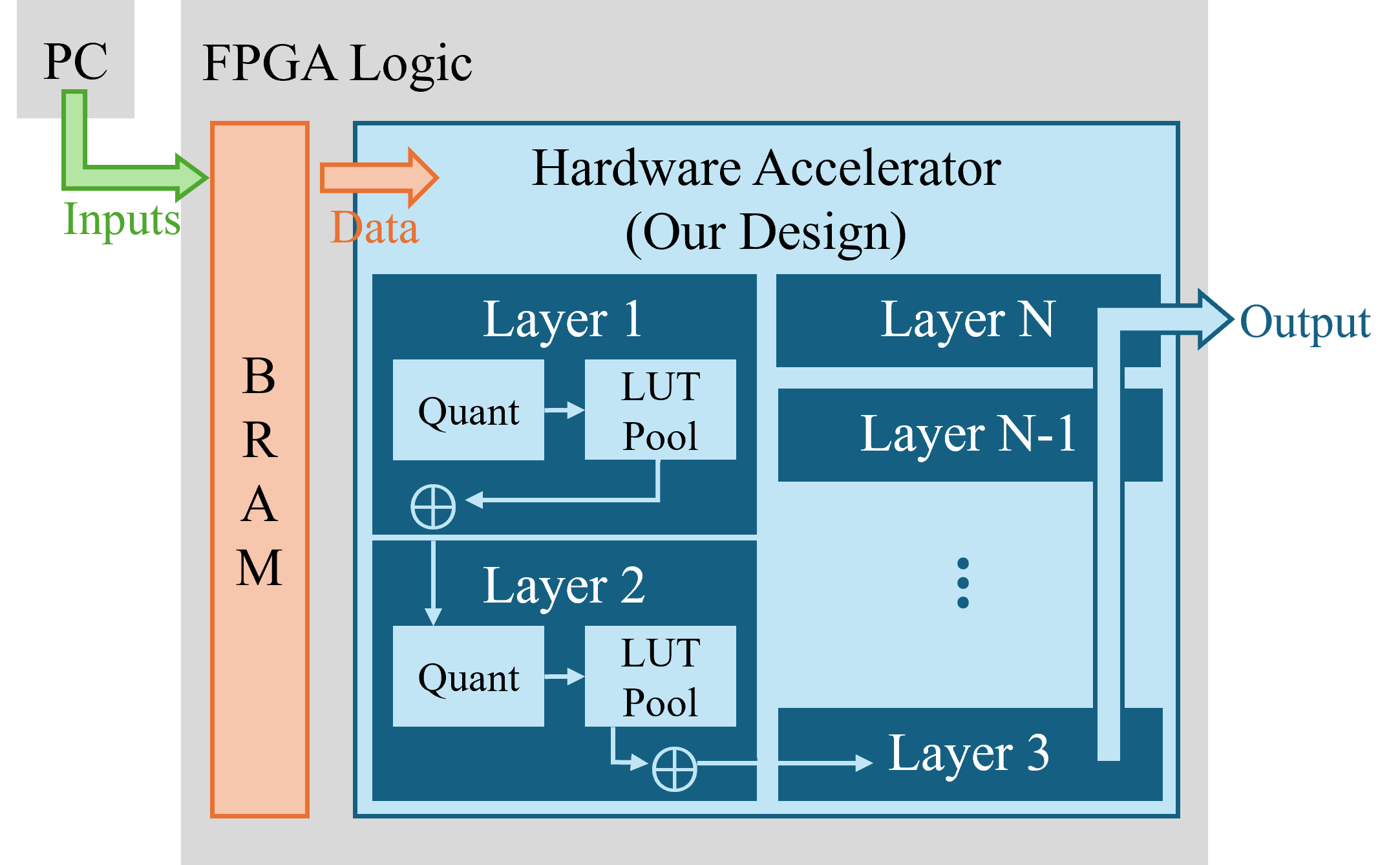}
    \caption{Overall hardware architecture of the accelerator.}
    \label{fig:overall}
\end{figure}

\begin{figure}[t]
    \centering
    \includegraphics[width=\columnwidth]{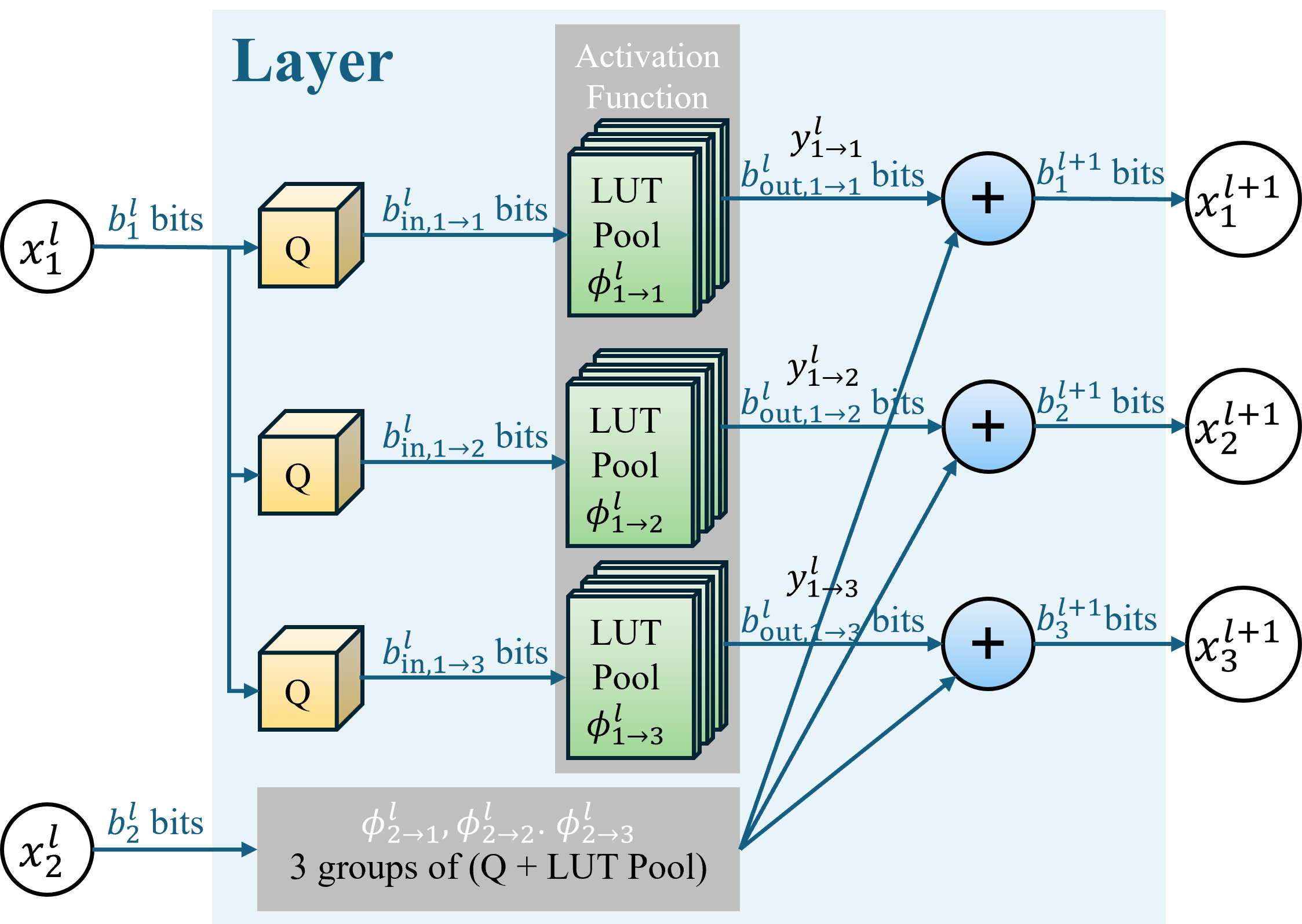}
   \caption{A single layer \(l\) of our proposed architecture with \(n_l = 2\) neurons and \(n_{l+1} = 3\) neurons in the subsequent layer.}
    \label{fig:act_func}
\end{figure}

The complete system design is illustrated in Figure~\ref{fig:overall}. The overall hardware accelerator consists of \(N\) layers, each following the architecture shown in Figure~\ref{fig:act_func}. In the figure, each \(x\) represents a neuron while each $Q$ denotes a quantization block.

In our hardware design process, key parameters can be configured, such as the number of neurons per layer and the bit-widths of the input and output of the LUT tables and quantization blocks. Hardware modules are instantiated on the basis of these parameters, enabling flexibility to support different models and quantization schemes.

To streamline deployment, we developed a fully automated toolchain that combines Python scripting with Vivado TCL automation. This toolchain automatically updates model instantiation scripts, generates per-edge LUT initialization files, configures quantization parameters, and constructs the corresponding Vivado project for synthesis and implementation. The overhead for generating model-specific files and runnable scripts generally takes 5–10 seconds depending on the host CPU. Modifying the model requires full resynthesis and bitstream generation, which takes approximately 6–10 hours. This one-time compilation overhead is acceptable for edge-device deployment scenarios where the model remains fixed over extended durations—for use cases such as temporal prediction—where frequent reprogramming is unnecessary.

\subsection{Quantization Block}    
After the accumulation in the previous layer \(l{-}1\), the \(i^{\text{th}}\) output neuron, denoted as \(x_i^l\), is represented using the precision level determined by the smallest step size in that layer.  
To align this precision with the input quantization required by the LUT pool in the current layer \(l\), a conversion is necessary.  

This task is handled by the \textit{quantization block}, which performs the precision adjustment described in Section~\ref{sec: quant_act_data}.  
Specifically, it converts the bitwidth of \(x_i^l\) from its original representation \(b_i^l\) to the target bitwidth \(b^l_{\text{in}, i \rightarrow j}\), ensuring compatibility with the input format expected by the LUT pool, where the activation function \(\phi^l_{i \rightarrow j}\) is applied.

Each input neuron value is scaled by an edge-specific factor \(\bar{\alpha}^{l}_{i \rightarrow j}\), represented as a fixed-point binary number to enable efficient hardware implementation.  
To correctly interpret this fixed-point value, the position of the binary (radix) point—which determines how many bits are allocated to the fractional part—must also be stored.  
The precision conversion is then carried out as illustrated in Equation~\ref{eq:precision_conv}.

Two cycles are allocated for the quantization block: one for the fixed-point multiplication and one for the rightward bit-shift operation.

For the edge connecting the \(i^{\text{th}}\) neuron in layer \(l\) to the \(j^{\text{th}}\) neuron in the next layer, the quantization block performs the following operation:
\begin{equation}
    \text{Q}_{i\rightarrow j}(x^{l}_{i}) = \bar{\alpha}^{l}_{i \rightarrow j} \cdot x^{l}_{i}
\end{equation}
 
\subsection{LUT Pool}
A LUT pool functions as a mapping tool. 
During the compilation process, each activation function $y^l_{i\rightarrow j}=\phi^l_{i\rightarrow j}(x_i^l)$ is quantized using either the global or fine-grained quantization method, with the software determining the optimal step size and range to trim. 

The input and output bitwidths \(b^l_{\text{in},i \rightarrow j}\) and \(b^l_{o<i \rightarrow j>}\) of the LUT pool for the activation function \(\phi^l_{i \rightarrow j}\) is determined by software quantization scheme according to Equation~\ref{eq:bo}. 
The composition of a LUT pool depends on its input and bitwidths \(b^l_{\text{in},i \rightarrow j}\) and \(b^l_{\text{out},i \rightarrow j}\).
The size of an individual LUT is constrained by the hardware architecture, where the fundamental LUT block in FPGA fabrics typically supports input sizes ranging from 1 to 6 bits. 
Therefore, the LUT size is set to \(\min(b^l_{\text{in},i \rightarrow j}, 6)\), ensuring compatibility with the FPGA's architecture while efficiently utilizing available logic resources. 
To construct a complete LUT pool that accommodates the required bitwidth, multiple fundamental LUT blocks may be needed. 
When the desired input bitwidth exceeds 6, the implementation must partition the input space and use multiple LUT blocks to cover all input combinations.  
Additionally, each output bit requires its own corresponding LUT mapping.  
The total number of fundamental LUT blocks required in the pool is given by
\[
b^l_{\text{out},i \rightarrow j} \cdot 2^{\max(0, b^l_{\text{in},i \rightarrow j} - 6)},
\]
where the term \(b^l_{\text{out},i \rightarrow j}\) accounts for the number of output bits, and the exponential term \(2^{\max(0, b^l_{\text{in},i \rightarrow j} - 6)}\) reflects the number of input partitions needed when the input bitwidth exceeds the 6-bit capacity of a single LUT.

The quantized mapping values are recorded and incorporated into the LUT initialization during compilation and synthesized into the FPGA's bitstream.

\subsection{Accumulator}
According to the model architecture, for each output neuron of a layer, the value after activation function for all input neurons must be added together to obtain the output neuron value.
% Since the LUT pool outputs that contribute to a single neuron value for the next layer have different quantization steps and ranges, they must be standardized before addition.
% This standardization is achieved by recording the scaling factor and offset for each LUT pool output. 
Since the LUT pool outputs that contribute to a single neuron value for the next layer have different ranges, they must be standardized before addition.

% Let \( s_{i \rightarrow j}^l \) represents the quantization step size for the LUT pool output \(y_{i \rightarrow j}^l\). The scaling factor \(\lambda_{i \rightarrow j}^l\) is calculated as the number of times \( s_{i \rightarrow j}^l \) exceeds the smallest quantization step among all LUT pool outputs connected to a single neuron in the next layer, defined as \(\min_{i \in [1, n^l]} s_{i \rightarrow j}^l\), so that \(\lambda_{i \rightarrow j}^l = \frac{s_{i \rightarrow j}^l}{\min_{i \in [1, n^l]} s_{i \rightarrow j}^l}\).

% The scaling factor \(\lambda_{i \rightarrow j}^l\) is calculated as the number of times \( s_{i \rightarrow j}^l \) exceeds the smallest quantization step among all LUT pool outputs connected to a single neuron in the next layer, defined as \(\min_{i \in [1, n^l]} s_{i \rightarrow j}^l\), so that \(\lambda_{i \rightarrow j}^l = \frac{s_{i \rightarrow j}^l}{\min_{i \in [1, n^l]} s_{i \rightarrow j}^l}\).

% The offset \(\delta_{i \rightarrow j}^l\), is determined by the difference in lower range between \(p_{i \rightarrow j}^l\) and the output with the smallest quantization step size \(\min_{i \in [1, n^l]} s_{i \rightarrow j}^l\).

Let \( s^l \) denote the shared quantization step size for all LUT pool outputs \( \phi_{i \rightarrow j}^l(\cdot) \) that contribute to the same output neuron \( j \) in layer \( l+1 \). Since the step size is uniform, there is no need to rescale the outputs prior to aggregation. However, as discussed in Section~\ref{sec: quant_fine_grain}, it is necessary to add the first-level offset of each activation function \( \phi_{i \rightarrow j}^l(\cdot) \) before aggregation, in order to compute the output of the next-layer neuron. This offset is given by \( \min(\phi_{i \rightarrow j}^l(\cdot)) \).

The standardized value of \( y_{i \rightarrow j}^l = \phi_{i \rightarrow j}^l(x_i^l) \) is then computed as:
\begin{equation}
    \text{standardize}(y_{i \rightarrow j}^l) = y_{i \rightarrow j}^l + \delta_{i \rightarrow j}^l,
    \label{eq:standardization}
\end{equation}
where \( \delta_{i \rightarrow j}^l = \min(\phi_{i \rightarrow j}^l(\cdot)) \) is the corresponding offset, expressed in units of the quantization step size \( s^l \).

It is important to note that the number of bits required to represent the standardized values must be increased to accommodate the largest dynamic range among all activation functions that contribute to the same output neuron in this layer.

In practical implementations, the standardization process can be integrated with the quantization block of the next layer, simplifying computations as follows. Here, \(i, j, k\) are the neuron indexes in layers $l, l+1, l+2$ respectively.
\begin{align}
   \text{Q}_{j\rightarrow k}({x}_j^{l+1}) 
     & = \text{Q}_{j\rightarrow k}(\sum_i{\text{standardize}(y_{i \rightarrow j}^l}))
   \\& = \bar{\alpha}^{l+1}_{j \rightarrow k} \cdot \sum_i{(y_{i \rightarrow j}^l + \delta_{i \rightarrow j}^l)}
   \\& = \sum_i\bar{\alpha}^{l+1}_{j \rightarrow k}\cdot y_{i \rightarrow j}^l + (\bar{\alpha}^{l+1}_{j \rightarrow k} \cdot \delta_{i \rightarrow j}^l)
\end{align}

This means that for each edge in the model, we need to precompute and store both the scaling factor \( \bar{\alpha}^{l+1}_{j \rightarrow k} \) and the corresponding offset term \( \bar{\alpha}^{l+1}_{j \rightarrow k} \cdot \delta_{i \rightarrow j}^l \).

After standardization, the outputs from the LUT pool are summed for each output neuron, as defined in Equation~\ref{eq:layer_definition}.

To fully leverage the capabilities of the FPGA, the accumulator is implemented using DSP slices. 
These slices are dedicated hardware blocks optimized for high-speed arithmetic operations, making them ideal for repeated additions involved in accumulation. 
The accumulation process is distributed across multiple cycles to balance resource utilization, which is especially important when working with large-scale neural networks that have a high number of incoming connections per neuron.
The number of cycles required for the whole accumulation depends on the shape of the model, specifically, the number of neurons in each layer and their connectivity. 
To handle this variability, handshake signals are implemented. These signals facilitate communication between hardware components, ensuring proper synchronization and preventing data inconsistencies.
% \clearpage
\section{Experimental Results}
\label{sec: result}

\subsection{Setup}
Two benchmarks are used to evaluate the performance of our proposed loop-up based acceleration platform.

The first benchmark is the spherical harmonics function, implemented as \texttt{ scipy.special.sph\_harm} in the Python SciPy library, using parameters \(m=0\) and \(n=2\). For this function, two KAN models are used to perform regression, respectively predicting the real and imaginary parts of the output from the input azimuthal coordinate \( \theta \) and the polar coordinate \( \phi \).
We use the KAN architecture proposed in \cite{kan_liu}, attaining a root mean square error (RMSE) of $1.874\times10^{-5}$.

The second is the MNIST dataset of handwritten digits, which has a training set of 60,000 examples and a test set of 10,000 examples~\cite{mnist}, which entails a classification task with \(28 \times 28 = 784 \) input dimensions and 10 output classes.
We train a KAN with dimensions (784, 64, 32, 10), achieving a classification accuracy of 98.52\%.

% The third is the MIT-BIH Arrhythmia dataset, which contains two-channel ambulatory ECG recordings~\cite{ecg}. Each input feature consists of a window of [-90, 90] samples centered on the R-peak, and hence has a dimension of 180. We perform binary classification on this dataset, categorizing the input signals into normal and abnormal classes.
% For this task, we train a KAN with dimensions (180, 512, 256, 2), achieving a classification accuracy of 91.05\%.

\subsection{Resource Savings from Quantization}
In lookup-based computing, the power draw is generally related to the number of LUT resources used.
Hence, we evaluate the impact of our quantization scheme on the number of logical LUT-4s required, obtained from software compilation.

\begin{table}[h]
\centering
\resizebox{\columnwidth}{!}{
    \begin{tabular}{lccccrrr}
    \hline
    \multirow{2}{*}{\textbf{Task}} & \multicolumn{2}{c}{\textbf{Output Bits}} & \multicolumn{2}{c}{\textbf{Input Bits}}  & \multirow{2}{*}{\textbf{LUT-4s}} & \multirow{2}{*}{\textbf{LUT-6s}} & \multirow{2}{*}{\textbf{Savings}} \\ 
    & Max & Mean & Max & Mean \\ \hline
    \multirow{3}{5.5em}{sph\_harm (16-bit input)} & 22 & - & 16 & - & 1,351,680 & 337920 & Baseline \\
    & 22 & 15.67 & 16 & - & 962,560 & 240640 & 28.78\% \\
    & 22 & 15.67 & 16 & 7.00 & 389,138 & 97285 & 71.21\% \\
    \hline
    \multirow{3}{5.5em}{sph\_harm (18-bit input)} & 22 & - & 18 & - & 5,406,720 & 1351680 & Baseline \\
    & 22 & 15.67 & 18 & - & 3,850,240 & 962560 & 28.78\% \\
    & 22 & 15.67 & 18 & 6.66 & 1,507,346 & 376837 & 72.12\% \\ 
    \hline
    \multirow{3}{*}{MNIST} & 5 & - & 4 & - & 262,720 & 65680 & Baseline \\
    & 5 & 2.29 & 4 & - & 120,539 & 30135 & 54.11\% \\
    & 5 & 2.29 & 4 & 3.77 & 113,484 & 28371 & 56.80\% \\
    \hline
    \end{tabular}
}
\vspace{0.2cm}
\caption{Resource use and percentage savings from quantization. Each task has three rows presenting (top) global quantization only, (middle) fine-grained outputs, and (bottom) fine-grained input and outputs.}
\label{tab:results-quant}
\end{table}

The LUT-4 resource demands and percentage savings are shown in Table \ref{tab:results-quant}.
For each task, we assess resource demands after applying applying (top row) global quantization only, (middle row) fine-grained quantization on outputs, and (bottom row) fine-grained quantization for inputs and outputs.

In the case of spherical harmonics, fine-grained quantization brings significant resource savings of more than 70\%.
Moreover, input quantization from $16 \rightarrow 7.00$ and $18 \rightarrow 6.66$ mean bits is especially impactful due to the exponential scaling with input bit-width, shown in Equation \ref{eq:lut-scaling}.

In the case of MNIST, fine-grained quantization also brings up to 56.80\% resource savings, but the allowance for fine-grained input quantization is lower (only $4 \rightarrow 3.77$).
Our observations suggest that fine-grained quantization will be most suitable for tasks that require high precision in several important outlier functions, such as in the case of spherical harmonics with 16- to 18-bit inputs and 22-bit outputs.

\subsection{Architectural Performance and Energy}
Our proposed architecture is implemented on the AMD Xilinx Virtex UltraScale+ XCVU13P FPGA with a clock frequency equal to 100MHz. We compare its performance in terms of accuracy, computation latency, and energy consumption against two other edge platforms: the Raspberry Pi 4 (8GB, CPU) and the NVIDIA Jetson AGX Orin (64GB, GPU). The results are shown in Table~\ref{tab:results}.

\begin{table*}[!h]
\centering
\begin{tabular}{lllcrr}
\hline
\rowcolor[gray]{0.9}
\textbf{Task} & \textbf{Model} & \textbf{Device} & \textbf{RMSE} & \textbf{Latency (us)} & \textbf{Energy (J)} \\ \hline

\multirow{6}{*}{\shortstack[l]{Spherical \\ Harmonics \\ (m=2, n=0)}} & SciPy & Edge CPU & 0 (FP) & 51.90 & $5.55 \times 10^{-5}$ \\ \cline{2-6} 
& \multirow{5}{*}{\shortstack[l]{KAN \\ (2, 5, 1)}} & Edge CPU & $1.874 \times 10^{-5}$ & 68.80 & $1.87 \times 10^{-4}$\\ \cline{3-6} 
& & Edge GPU & $1.874 \times 10^{-5}$ & 1.60 & $2.75 \times 10^{-5}$ \\ \cline{3-6} 
& & \multirow{2}{*}{\shortstack[l]{Our Design \\ (18-bit input, fine-grained)}} &  \multirow{2}{*}{\shortstack[l]{$1.902 \times 10^{-5}$}} & \multirow{2}{*}{\shortstack[l]{0.26}} & \multirow{2}{*}{\shortstack[l]{$3.54 \times 10^{-7}$}} \\ 
& & & & & \\ \cline{3-6} 
 
& & \multirow{2}{*}{\shortstack[l]{Our Design \\ (16-bit input, fine-grained)}} &  \multirow{2}{*}{\shortstack[l]{$3.003 \times 10^{-5}$}} &  \multirow{2}{*}{\shortstack[l]{0.26}} &  \multirow{2}{*}{\shortstack[l]{$8.06 \times 10^{-9}$}} \\ 
& & & & & \\ \hline

\rowcolor[gray]{0.9}
\textbf{Task} & \textbf{Model} & \textbf{Device} & \textbf{Accuracy (\%)} & \textbf{Latency (us)} & \textbf{Energy (J)} \\ \hline

\multirow{6}{*}{MNIST} & \multirow{2}{*}{\shortstack[l]{MLP \\ (784, 64, 32, 10)}} & Edge CPU & 98.38 & 344.80 & $6.85 \times 10^{-4}$ \\ \cline{3-6} 
 &  & Edge GPU & 98.38 & 73.29 & $3.35 \times 10^{-4}$ \\ \cline{2-6}
 &  \multirow{4}{*}{\shortstack[l]{KAN \\ (784, 64, 32, 10)}} & Edge CPU & 98.52 & 45,552.80 & $1.13 \times 10^{-1}$ \\ \cline{3-6}
  &   & Edge GPU & 98.52 & 984.60 & $1.94 \times 10^{-2}$ \\ \cline{3-6}
 &  &  \multirow{2}{*}{\shortstack[l]{Our Design \\ (4-bit input, fine-grained)}}& \multirow{2}{*}{\shortstack[l]{98.02}} & \multirow{2}{*}{\shortstack[l]{4.74}} &  \multirow{2}{*}{\shortstack[l]{$2.90 \times 10^{-6}$}}\\ 
 & & & & & \\\hline
\end{tabular}
\vspace{0.2cm}
\caption{Comparison of performance across different tasks and hardware platforms.}
\label{tab:results}
\end{table*}

\textbf{Accuracy} For the spherical harmonics function benchmark, accuracy is measured using RMSE, while for MNIST, it is evaluated as the classification accuracy. As shown in Table~\ref{tab:results}, our quantization scheme ensures that the accuracy of the proposed model remains within the same order of magnitude as the floating-point KAN and MLP implementations, demonstrating the robustness of the approach.

\textbf{Latency} It can be observed that our architecture accelerates KAN models, outperforming both the edge CPU Raspberry Pi 4 and GPU NVIDIA Jetson AGX Orin. The speedup is more pronounced for larger models. For the spherical harmonics function benchmark, it achieves a one-order-of-magnitude ($10\times$) speedup over the NVIDIA Jetson AGX Orin, while for the MNIST benchmark, it reaches a two-order-of-magnitude ($100\times$) speedup over the NVIDIA Jetson AGX Orin. Moreover, since the design is fully pipelined and produces one output per cycle, it achieves high throughput.

% \textcolor{red}{Only one sentence"Across all tasks, our design shows lower energy consumption compared to both the edge CPU and the
% GPU." is used to concretely discuss the results. Could you discuss/analysis more on the results shown in the table? Additionally, since the measurements were taken using a metered power supply for the Raspberry Pi 4 and NVIDIA Jetson AGX Orin. could you show some photo and mark the metered power supply to enhance the credibility?}

\textbf{Energy Consumption} The energy consumption reported in the last column of Table~\ref{tab:results} represents the dynamic energy, i.e., the energy required to process a single test sample, excluding the baseline power required to keep the hardware running. The dynamic energy for a single test sample was computed by multiplying the latency by the dynamic power consumption. Measurements were taken using a metered power supply for the Raspberry Pi 4 and NVIDIA Jetson AGX Orin. The power readings are averaged over the duration of the experiment, while the static power is measured by exerting no load for at least 5 seconds. The dynamic energy consumption for a single test sample is calculated by multiplying the latency by the dynamic power consumption. Across all tasks, our design shows lower energy consumption compared to both the edge CPU and the GPU. 

For the Spherical Harmonics task, the 18-bit and 16-bit fine-grained implementations achieve energy reductions of approximately \(5 \times 10^2\) and \(2 \times 10^4\), respectively, compared to the edge CPU, and about \(7.7 \times 10^1\) and \(3 \times 10^3\) compared to the edge GPU. 
In the MNIST task, our 4-bit fine-grained design achieves an energy consumption of just \(2.90 \times 10^{-6}\)~J, which is nearly \(4 \times 10^4\) times lower than that of the edge CPU and more than \(10^2\) times lower than that of the edge GPU.

These substantial reductions stem from two key factors: reduced execution time and lower dynamic power.
The reduced execution time has been discussed earlier.
In addition, the use of fine-grained quantization and lightweight, lookup-based operations—instead of general-purpose arithmetic—reduces switching activity, which is the primary contributor to dynamic power in digital circuits.

\section{Conclusion}
\label{sec: conclusion}

In conclusion, this work introduces an innovative lookup-based architecture designed to accelerate neural networks with complex, learnable activation functions, such as KANs. By leveraging the high representational power of lookup tables and implementing a fine-grained quantization scheme, the proposed solution significantly reduces both latency and energy consumption compared to standard CPU and GPU implementations, while preserving a high level of accuracy. The hardware design is fully automated and adaptable to different quantization levels and model shapes, taking full advantage of the reconfigurable capabilities of FPGAs. A current limitation of this work lies in the relatively small scale of the evaluated model, which makes it challenging to fully assess the scalability of the proposed architecture.

Future research directions include:

\begin{itemize}

    \item \textbf{Addressing scalability challenges}: By identifying similarities among activation functions, multiple functions can share a common lookup table. This approach reduces the overall resource footprint, enabling resource-constrained FPGAs to accommodate larger neural networks.

    \item \textbf{Optimizing and automating the design space exploration process}: Future work includes developing a selection mechanism between conventional arithmetic cores and lookup-based units based on model dimensions, hardware constraints, and resource availability to further improve efficiency. Integrating this mechanism into the workflow could further enhance hardware efficiency.

\end{itemize}
\bibliographystyle{IEEEtran}
\bibliography{references}

% Generated by IEEEtran.bst, version: 1.14 (2015/08/26)
\begin{thebibliography}{10}
\providecommand{\url}[1]{#1}
\csname url@samestyle\endcsname
\providecommand{\newblock}{\relax}
\providecommand{\bibinfo}[2]{#2}
\providecommand{\BIBentrySTDinterwordspacing}{\spaceskip=0pt\relax}
\providecommand{\BIBentryALTinterwordstretchfactor}{4}
\providecommand{\BIBentryALTinterwordspacing}{\spaceskip=\fontdimen2\font plus
\BIBentryALTinterwordstretchfactor\fontdimen3\font minus \fontdimen4\font\relax}
\providecommand{\BIBforeignlanguage}[2]{{%
\expandafter\ifx\csname l@#1\endcsname\relax
\typeout{** WARNING: IEEEtran.bst: No hyphenation pattern has been}%
\typeout{** loaded for the language `#1'. Using the pattern for}%
\typeout{** the default language instead.}%
\else
\language=\csname l@#1\endcsname
\fi
#2}}
\providecommand{\BIBdecl}{\relax}
\BIBdecl

\bibitem{dnn_trainable_act}
H.~Chung, S.~J. Lee, and J.~G. Park, ``Deep neural network using trainable activation functions,'' in \emph{2016 International Joint Conference on Neural Networks (IJCNN)}, 2016, pp. 348--352.

\bibitem{relu}
\BIBentryALTinterwordspacing
X.~Glorot, A.~Bordes, and Y.~Bengio, ``Deep sparse rectifier neural networks,'' in \emph{Proceedings of the Fourteenth International Conference on Artificial Intelligence and Statistics}, ser. Proceedings of Machine Learning Research, G.~Gordon, D.~Dunson, and M.~Dudík, Eds., vol.~15.\hskip 1em plus 0.5em minus 0.4em\relax Fort Lauderdale, FL, USA: PMLR, 11--13 Apr 2011, pp. 315--323. [Online]. Available: \url{https://proceedings.mlr.press/v15/glorot11a.html}
\BIBentrySTDinterwordspacing

\bibitem{leaky_relu}
A.~L. Maas, A.~Y. Hannun, A.~Y. Ng \emph{et~al.}, ``Rectifier nonlinearities improve neural network acoustic models,'' in \emph{Proc. icml}, vol.~30, no.~1.\hskip 1em plus 0.5em minus 0.4em\relax Atlanta, GA, 2013, p.~3.

\bibitem{trainable_act_func_survey}
\BIBentryALTinterwordspacing
A.~Apicella, F.~Donnarumma, F.~Isgrò, and R.~Prevete, ``A survey on modern trainable activation functions,'' \emph{Neural Networks}, vol. 138, pp. 14--32, 2021. [Online]. Available: \url{https://www.sciencedirect.com/science/article/pii/S0893608021000344}
\BIBentrySTDinterwordspacing

\bibitem{s_relu}
X.~Jin, C.~Xu, J.~Feng, Y.~Wei, J.~Xiong, and S.~Yan, ``Deep learning with s-shaped rectified linear activation units,'' in \emph{Proceedings of the AAAI conference on artificial intelligence}, vol.~30, no.~1, 2016.

\bibitem{adapt_act_func}
S.~Qian, H.~Liu, C.~Liu, S.~Wu, and H.~San~Wong, ``Adaptive activation functions in convolutional neural networks,'' \emph{Neurocomputing}, vol. 272, pp. 204--212, 2018.

\bibitem{kan_liu}
\BIBentryALTinterwordspacing
Z.~Liu, Y.~Wang, S.~Vaidya, F.~Ruehle, J.~Halverson, M.~Soljačić, T.~Y. Hou, and M.~Tegmark, ``Kan: Kolmogorov-arnold networks,'' 2025. [Online]. Available: \url{https://arxiv.org/abs/2404.19756}
\BIBentrySTDinterwordspacing

\bibitem{kan_liu_2}
\BIBentryALTinterwordspacing
Z.~Liu, P.~Ma, Y.~Wang, W.~Matusik, and M.~Tegmark, ``Kan 2.0: Kolmogorov-arnold networks meet science,'' 2024. [Online]. Available: \url{https://arxiv.org/abs/2408.10205}
\BIBentrySTDinterwordspacing

\bibitem{ma_bert}
\BIBentryALTinterwordspacing
N.~W. Ming, Z.~Wang, C.~Liu, R.~S.~M. Goh, and T.~Luo, ``Ma-bert: Towards matrix arithmetic-only bert inference by eliminating complex non-linear functions,'' in \emph{International Conference on Learning Representations}, 2023. [Online]. Available: \url{https://api.semanticscholar.org/CorpusID:259298534}
\BIBentrySTDinterwordspacing

\bibitem{finn}
\BIBentryALTinterwordspacing
Y.~Umuroglu, N.~J. Fraser, G.~Gambardella, M.~Blott, P.~Leong, M.~Jahre, and K.~Vissers, ``Finn: A framework for fast, scalable binarized neural network inference,'' in \emph{Proceedings of the 2017 ACM/SIGDA International Symposium on Field-Programmable Gate Arrays}, ser. FPGA '17.\hskip 1em plus 0.5em minus 0.4em\relax New York, NY, USA: Association for Computing Machinery, 2017, p. 65–74. [Online]. Available: \url{https://doi.org/10.1145/3020078.3021744}
\BIBentrySTDinterwordspacing

\bibitem{fpga}
\BIBentryALTinterwordspacing
E.~Nurvitadhi, G.~Venkatesh, J.~Sim, D.~Marr, R.~Huang, J.~Ong Gee~Hock, Y.~T. Liew, K.~Srivatsan, D.~Moss, S.~Subhaschandra, and G.~Boudoukh, ``Can fpgas beat gpus in accelerating next-generation deep neural networks?'' in \emph{Proceedings of the 2017 ACM/SIGDA International Symposium on Field-Programmable Gate Arrays}, ser. FPGA '17.\hskip 1em plus 0.5em minus 0.4em\relax New York, NY, USA: Association for Computing Machinery, 2017, p. 5–14. [Online]. Available: \url{https://doi.org/10.1145/3020078.3021740}
\BIBentrySTDinterwordspacing

\bibitem{dsp_acc}
D.~Wang, K.~Xu, J.~Guo, and S.~Ghiasi, ``Dsp-efficient hardware acceleration of convolutional neural network inference on fpgas,'' \emph{IEEE Transactions on Computer-Aided Design of Integrated Circuits and Systems}, vol.~39, no.~12, pp. 4867--4880, 2020.

\bibitem{tlmac}
\BIBentryALTinterwordspacing
D.~Gerlinghoff, B.~C.~M. Choong, R.~S.~M. Goh, W.-F. Wong, and T.~Luo, ``Table-lookup mac: Scalable processing of quantised neural networks in fpga soft logic,'' in \emph{Proceedings of the 2024 ACM/SIGDA International Symposium on Field Programmable Gate Arrays}, ser. FPGA '24.\hskip 1em plus 0.5em minus 0.4em\relax New York, NY, USA: Association for Computing Machinery, 2024, p. 235–245. [Online]. Available: \url{https://doi.org/10.1145/3626202.3637576}
\BIBentrySTDinterwordspacing

\bibitem{cardinaux2020iteratively}
F.~Cardinaux, S.~Uhlich, K.~Yoshiyama, J.~A. Garc{\'\i}a, L.~Mauch, S.~Tiedemann, T.~Kemp, and A.~Nakamura, ``Iteratively training look-up tables for network quantization,'' \emph{IEEE Journal of Selected Topics in Signal Processing}, vol.~14, no.~4, pp. 860--870, 2020.

\bibitem{wang2022learnable}
L.~Wang, X.~Dong, Y.~Wang, L.~Liu, W.~An, and Y.~Guo, ``Learnable lookup table for neural network quantization,'' in \emph{Proceedings of the IEEE/CVF conference on computer vision and pattern recognition}, 2022, pp. 12\,423--12\,433.

\bibitem{guo2024fast}
H.~Guo, W.~Brandon, R.~Cholakov, J.~Ragan-Kelley, E.~P. Xing, and Y.~Kim, ``Fast matrix multiplications for lookup table-quantized llms,'' \emph{arXiv preprint arXiv:2407.10960}, 2024.

\bibitem{vasyltsov2021efficient}
I.~Vasyltsov and W.~Chang, ``Efficient softmax approximation for deep neural networks with attention mechanism,'' \emph{arXiv preprint arXiv:2111.10770}, 2021.

\bibitem{chen2022approximate}
K.~Chen, Y.~Gao, H.~Waris, W.~Liu, and F.~Lombardi, ``Approximate softmax functions for energy-efficient deep neural networks,'' \emph{IEEE Transactions on Very Large Scale Integration (VLSI) Systems}, vol.~31, no.~1, pp. 4--16, 2022.

\bibitem{yu2022nn}
J.~Yu, J.~Park, S.~Park, M.~Kim, S.~Lee, D.~H. Lee, and J.~Choi, ``Nn-lut: neural approximation of non-linear operations for efficient transformer inference,'' in \emph{Proceedings of the 59th ACM/IEEE Design Automation Conference}, 2022, pp. 577--582.

\bibitem{wang2019lutnet}
E.~Wang, J.~J. Davis, P.~Y. Cheung, and G.~A. Constantinides, ``Lutnet: Rethinking inference in fpga soft logic,'' in \emph{2019 IEEE 27th Annual International Symposium on Field-Programmable Custom Computing Machines (FCCM)}.\hskip 1em plus 0.5em minus 0.4em\relax IEEE, 2019, pp. 26--34.

\bibitem{wang2022logic}
E.~Wang, J.~J. Davis, G.-I. Stavrou, P.~Y. Cheung, G.~A. Constantinides, and M.~Abdelfattah, ``Logic shrinkage: Learned fpga netlist sparsity for efficient neural network inference,'' in \emph{Proceedings of the 2022 ACM/SIGDA International Symposium on Field-Programmable Gate Arrays}, 2022, pp. 101--111.

\bibitem{umuroglu2020logicnets}
Y.~Umuroglu, Y.~Akhauri, N.~J. Fraser, and M.~Blott, ``Logicnets: Co-designed neural networks and circuits for extreme-throughput applications,'' in \emph{2020 30th International Conference on Field-Programmable Logic and Applications (FPL)}.\hskip 1em plus 0.5em minus 0.4em\relax IEEE, 2020, pp. 291--297.

\bibitem{andronic2023polylut}
M.~Andronic and G.~A. Constantinides, ``Polylut: learning piecewise polynomials for ultra-low latency fpga lut-based inference,'' in \emph{2023 International Conference on Field Programmable Technology (ICFPT)}.\hskip 1em plus 0.5em minus 0.4em\relax IEEE, 2023, pp. 60--68.

\bibitem{park2022lut}
G.~Park, B.~Park, M.~Kim, S.~Lee, J.~Kim, B.~Kwon, S.~J. Kwon, B.~Kim, Y.~Lee, and D.~Lee, ``Lut-gemm: Quantized matrix multiplication based on luts for efficient inference in large-scale generative language models,'' \emph{arXiv preprint arXiv:2206.09557}, 2022.

\bibitem{jin2022f8net}
Q.~Jin, J.~Ren, R.~Zhuang, S.~Hanumante, Z.~Li, Z.~Chen, Y.~Wang, K.~Yang, and S.~Tulyakov, ``F8net: Fixed-point 8-bit only multiplication for network quantization,'' \emph{arXiv preprint arXiv:2202.05239}, 2022.

\bibitem{mnist}
L.~Deng, ``The mnist database of handwritten digit images for machine learning research,'' \emph{IEEE Signal Processing Magazine}, vol.~29, no.~6, pp. 141--142, 2012.

\end{thebibliography}

\end{document}